\documentclass[iop,twocolumn]{aastex6}
\usepackage{amsmath}
\usepackage{gensymb}

\newcommand\cs{c_s}

\newcommand\Msun{\; {\rm M}_{\odot}}

\newcommand\kms{\; {\rm km}\;{\rm s}^{-1}}
\newcommand\pc{\;{\rm pc}}
\newcommand\eV{\;{\rm eV}}

\newcommand\kpc{\;{\rm kpc}}

\newcommand\freq{\kms\kpc^{-1}}

\newcommand\Myr{\;{\rm Myr}}
\newcommand\Gyr{\;{\rm Gyr}}

\newcommand\Surf{\Msun\;{\rm pc^{-2}}}

\newcommand\simgt{\lower.5ex\hbox{$\; \buildrel > \over \sim \;$}}
\newcommand\simlt{\lower.5ex\hbox{$\; \buildrel < \over \sim \;$}}
\newcommand{\RNum}[1]{\uppercase\expandafter{\romannumeral #1\relax}}

\def\lvplot{($l,v$) diagram}
\def\lvplots{($l,v$) diagrams}

\def\spose#1{\hbox to 0pt{#1\hss}}
\def\dt{\spose{\raise 1.0ex\hbox{\hskip2pt$\mathchar"201$}}}

\shorttitle{Soliton core in the Milky Way}
\shortauthors{Li, Shen \& Schive}

\begin{document}

\title{Testing the Prediction of Fuzzy Dark Matter Theory in the Milky Way Center}


\author{Zhi Li \altaffilmark{1}, 
Juntai Shen \altaffilmark{2,3,4,5}, and 
Hsi-Yu Schive \altaffilmark{6}}

\affil{
$^1$Tsung-Dao Lee Institute, Shanghai Jiao Tong University, Shanghai 200240, China; email: zli0804@sjtu.edu.cn\\
$^2$Department of Astronomy, School of Physics and Astronomy, Shanghai Jiao Tong University, Shanghai 200240, China; email: jtshen@sjtu.edu.cn\\
$^3$Shanghai Key Laboratory for Particle Physics and Cosmology, 200240, Shanghai, China\\
$^4$Shanghai Astronomical Observatory, Chinese Academy of Sciences, 80
Nandan Road, Shanghai 200030, China\\
$^5$College of Astronomy and Space Sciences, University of Chinese Academy
of Sciences, 19A Yuquan Road, Beijing 100049, China\\
$^6$Center for Theoretical Physics, National Taiwan University, Taipei 10617, Taiwan; email:hyschive@phys.ntu.edu.tw}


\begin{abstract}
The fuzzy dark matter model (FDM, also known as quantum wave dark matter model) argues that light bosons with a mass of $\sim10^{-22}\eV$ are a possible candidate for dark matter in the Universe. 
One of the most important predictions of FDM is the formation of a soliton core instead of a density cusp at the center of galaxies. If FDM is the correct theory of dark matter, then the predicted soliton core can help to form the Central Molecular Zone (CMZ) in the Milky Way. We present high-resolution hydrodynamical simulations of gas flow patterns to constrain the properties of the soliton core based on a realistic Milky Way potential. We find that a dense center is required to form a reasonable CMZ. The size and kinematics of the CMZ offer a relatively strong constraint on the inner enclosed mass profile of the Galaxy. If a soliton core is not considered, a compact nuclear bulge alone with a radially varying mass-to-light ratio can match the observed size and kinematics of the CMZ. A soliton core model with a mass of $\approx4.0\times10^8\Msun$ and a core radius of $\approx0.05\kpc$, together with a less massive nuclear bulge with a constant mass-to-light ratio, also agrees nicely with the current data. Such a FDM soliton core corresponds to a boson mass of $\sim2-7\times10^{-22}\eV$, which could be further constrained by the improved determination of the mass-to-light ratio in the Galactic center.

\end{abstract}


\section{Introduction}

Although the current cold dark matter (CDM) model successfully 
explains many problems on the large-scale structures of the universe
\citep[e.g.][]{fre_whi_12,bennet_etal_13}, 
some of its predictions on the galactic scales are still in tension 
with modern observations \citep[see the review by][]{bul_boy_17}. 
One of the famous discrepancies is the 
so-called ``cusp-core'' problem: 
dissipationless CDM simulations predicted a 
universal Navarro-Frenk-White (NFW) density profile for 
bound dark matter halos, with a cuspy density profile 
at central part \citep{navarr_etal_97}; 
while observations favour a flat cored profile in 
low surface brightness galaxies (LSB) and 
dwarf galaxies \citep{blok_10}. 
A shallower dark matter core might be created 
by considering the various baryonic effects in simulations 
\citep[e.g.][]{onorbe_etal_15,read_etal_19}.
However, it is still not clear how 
these ``sub-grid physics'' alter the 
structure formation in the lowest mass galaxies
\citep[e.g.][]{sanche_etal_12,weinber_etal_15}. 

On the other side, alternative dark matter models are 
being developed to solve the problems in a more 
self-consistent way. In  
recent years the Fuzzy Dark Matter model (FDM), 
also known as
``quantum wave dark matter'' model, is gaining more 
attention. In this scenario, 
the dark matter is composed of very 
light bosons with a particle mass of 
$\sim 10^{-22}\eV$ 
\citep[e.g.][]{hu_etal_00,mar_sil_14,schive_etal_14a}. 
Such bosons have a characteristic wavelength of 
$mv/\hbar\sim0.2\kpc$ assuming a velocity of $100\kms$, 
which helps to suppress the formation of small-scale structures. 
One of the most important predictions of FDM is the presence of a stable 
ground state soliton core due to the Bose-Einstein condensate 
at the central part of each dark matter halo. The core 
radius is usually comparable to the characteristic wavelength, 
and the halo transitions to a NFW profile within 
a few core radii according to recent numerical simulations 
\citep[e.g.][]{schive_etal_14b}. 
This may provide a plausible solution to the ``cusp-core'' 
problem in CDM model. One can easily deduce 
that a lower mass halo would host a larger soliton core due to 
the particles inside it having a smaller velocity and a 
larger characteristic wavelength. Therefore previous works 
\citep[e.g.][]{chen_etal_17} preferred to use dwarf galaxies 
to constrain the soliton core mass and size, and a
boson mass of $1-2\times10^{-22}\eV$ is favored 
in such studies. 

Naturally, one may wonder whether the soliton core exists in our 
Milky Way and if so, how we can detect its effects 
on the stars and gas. The Milky Way should be 
a unique laboratory for testing the 
existence of such a soliton core due to the 
unprecedented data quality we could achieve compared 
to external galaxies. 
However, one should be careful when applying 
the halo-core scaling relation found in previous FDM simulations 
\citep[e.g.][]{schive_etal_14b} to the Milky-Way like
galaxies to derive the soliton core properties. 
One reason is that previous FDM simulations focus mainly on 
the halo mass range of $10^9-10^{11}\Msun$. It is not clear 
whether the scaling relation between the host 
halo mass and the central soliton core properties 
can be extrapolated 
to more massive halos, such as the Milky Way  
with a virial halo mass of $\approx1.5\times10^{12}\Msun$ 
\citep[]{portail_etal_17,li_etal_19,pos_hel_19}. Another reason is 
that the halo-core 
relation could be interpreted as the 
specific kinetic energy of the soliton core 
equals that of the halo. However, the specific 
kinetic energy of the soliton core can be strongly affected 
by the baryons in the Galactic center, and therefore the halo-core 
scaling relation may need to be adjusted. 
\citep[][see also \S\ref{sec:pot_sc}]{bar_etal_19}.

Nevertheless, 
according to the scaling relation in \citet{schive_etal_14b}, 
the expected soliton mass 
and core radius are $1.44\times10^9\Msun$ and 
$0.16\kpc$ for the Milky Way, 
respectively, by assuming a boson mass of 
$1.0\times10^{-22}\eV$. It is also important to note that 
these two numbers should be interpreted as the 
maximum soliton core mass and the minimum soliton core radius,
due to the fact that the halo of the Milky Way 
may still be relaxing \citep{hui_etal_17}.

Observations on stellar and gas kinematics in the Galactic 
center have revealed some evidence for a central component with 
mass of $10^9\Msun$, although it is not clear whether this component 
is composed of baryons or dark matter 
\citep[see the discussions in Sec.V.B of][]{bar_etal_18}. 
\citet{portail_etal_17} and \citet{martin_etal_18} 
included such a component in their dynamical models to explain 
the high velocity dispersion of the stars 
in the Galactic bulge. Moreover, the Central 
Molecular Zone \citep[CMZ,][]{bally_etal_88} which 
is thought to be a molecular 
gas ring or disk orbiting the Galactic center, 
has a size of $\approx100\pc$ 
(Galactic longitude $|l|\la1.5\degree$) with a rotation 
velocity of $\approx100\kms$ \citep[e.g.][]{molina_etal_11,kruijs_etal_15,hensha_etal_16}. 
This implies an enclosed mass of 
$\approx2.3\times10^8\Msun$ inside $\approx100\pc$ by assuming circular 
motions. The formation of the CMZ is 
related to the bar-induced gas inflow due to the large-scale 
bi-symmetric gravitational torques 
\citep[e.g.][]{sorman_etal_15a,li_etal_16,ridley_etal_17}.  
In addition, the CMZ needs a dense center ($\sim1\%$ disk mass) to support 
its backbone $x_2$ orbits 
\citep{reg_teu_03,li_etal_17,sorman_etal_18}. Previous studies 
usually included a very compact stellar nuclear bulge to form the CMZ 
\citep[e.g.][]{baba_etal_10,li_etal_16,shin_etal_17,sorman_etal_18,armill_etal_19}, 
but the physical origin of this component is still not clear. 
The mass and size of the nuclear bulge 
are also not well-constrained, due both to dust extinction 
on the Galactic plane and to the uncertainties on the mass-to-light ratio of the 
nuclear bulge (see \S\ref{sec:pot_nb}). 
This gives us some freedom to add a soliton core in our 
previous Milky Way gas dynamical model 
\citep{li_etal_16} to see whether we could still 
reproduce the observed CMZ shape and kinematics, 
thus putting important constraints on the FDM theory. 

The purpose of the present paper is to use high resolution 
hydrodynamical simulations to determine the allowed mass 
and size of the soliton core in the Milky Way. 
The paper is organized as follows: in \S\ref{sec:setup} 
we describe our model setup and numerical methods; 
in \S\ref{sec:simulationresults} we 
present our simulation results; In \S\ref{sec:discussions} 
we discuss the possibility of the existence of such a soliton core 
together with its additional dynamical effects. 
The summary is 
presented in \S\ref{sec:summary}. 

\section{Model Setup}
\label{sec:setup}

\subsection{Gravitational Potential}
\label{sec:pot_general}

\subsubsection{Realistic Milky Way Potential}
\label{sec:pot_m2m}

\begin{figure}[!t]
\includegraphics[width=0.5\textwidth]{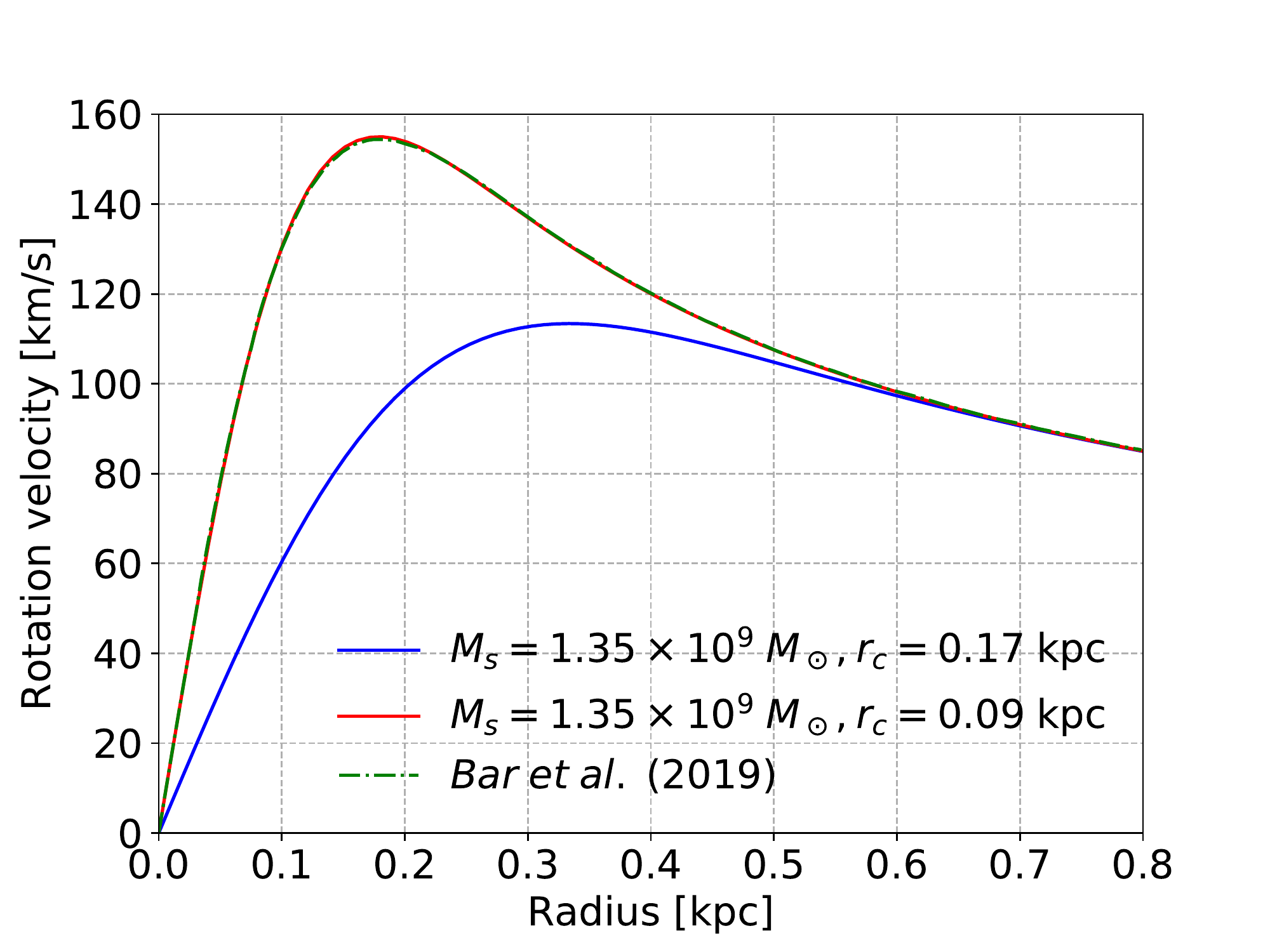}
\caption{The rotation curves of a soliton core with 
a fixed mass of $1.35\times10^9\Msun$ but different core radius 
calculated by Eq.~\ref{eq:scrotcurve}. 
In this plot we assume the virial mass of the Milky Way halo is 
$1.0\times10^{12}\Msun$ and
a boson particle mass of $1.0\times10^{-22}\eV$, same as in 
\citet{bar_etal_19}. The blue solid line shows the 
rotation curve of a self-gravitating soliton core with a core radius 
$r_c=0.17\kpc$, while the red dotted-dash line shows a soliton core that 
is compressed by a factor of 2. The compressed soliton core has the 
same total mass but a smaller core radius $r_c=0.09\kpc$. 
The green dotted-dashed curve shows the results from \citet{bar_etal_19} 
who obtained the compressed soliton core profile by directly 
solving Schr{\"o}dinger-Poisson equation in an 
external nuclear bulge potential. Their results are in good agreement
with Eq.~\ref{eq:scrotcurve}.
\label{fig:scrc}}
\vspace{0.2cm}
\end{figure}

The Galactic potential used in this work is based on the 
best-fit gas dynamical model of \citet{li_etal_16}, 
where they used the features in \lvplot\ to constrain the gas flow patterns in the 
Milky Way on large scales. The \lvplot\ shows 
the intensity of cold gas emission as a function 
of Galactic longitude $l$ and line-of-sight velocity $v$ \citep[e.g.][]{dame_etal_01}.
The gaseous features viewed from face-on 
appear as high-density ridges in the \lvplot, and these features 
are formed mainly due to the non-axisymmetric structures such as the 
Galactic stellar bar and spiral arms. 

The large-scale potential of the Milky Way has been 
studied extensively in recent works. 
\citet{portail_etal_15a} constructed a made-to-measure Milky Way 
model that matches the 3D density of red clump giants in the bulge 
region \citep[]{weg_ger_13} and the BRAVA kinematics \citep[]{kunder_etal_12} quite well.  
The axial ratios of the bar are (10:6.3:2.6) with the semi-major axis of $\approx5\kpc$.
The bar angle to the Sun-Galactic center line is $(27 \pm 2^{\circ})$. 
We adopt this model 
as the basis of the Galactic potential with a few minor 
modifications as in \citet{li_etal_16}.
The bar pattern speed is chosen to be $\Omega_b = 33\freq$, which 
places the corotation radius at $R=6.2\kpc$. 
In this study we remove the large-scale spiral potentials from \citet{li_etal_16} 
as they are only important for the outer gas features, which is not the main 
focus of this work.

On the other hand, the central mass distribution in our Galaxy is not well-constrained.
The stellar kinematics may only be reliable at higher latitudes ($b \ge 2^{\circ}$), 
due to the heavy dust contamination closer to the disk plane. The gas kinematics 
in the central disk region may not serve as a good indicator of mass either, due to 
its highly non-circular motions caused by the Galactic bar 
\citep[e.g.][]{fux_99b,sorman_etal_15b,sorman_etal_17a}. We therefore modify the 
central potential of the best-fit model in \citet{li_etal_16} to test the possible 
effects caused by a soliton core. We include the following two components, namely the
nuclear bulge and the soliton core, in the central part of the Milky Way potential. 

\subsubsection{Nuclear bulge}
\label{sec:pot_nb}

\citet{launha_etal_02} found hints of a stellar nuclear bulge component 
in the central $\sim200\pc$ of the Milky Way using COBE 
near-infrared image. The authors 
parameterized the nuclear bulge with the following equation:

\begin{equation}
\label{eq:nbprofile}
\begin{split}
\rho_{\rm{nb}}(R,z) & = 32.4~\Upsilon~\exp{ \{} \\
                    & {- 0.693 [(\frac{R}{R_{\rm{nb}}})^5 + (\frac{z}{0.045\kpc})^{1.4}]} \}~\Msun\pc^{-3},
\end{split}
\end{equation}

where $\Upsilon$ is the mass-to-light ratio, 
and $R_{\rm{nb}}$ is the scale radius of the nuclear bulge. 
\citet{launha_etal_02} used two components with different scale radii, 
$R_{\rm{nb,1}}=120\pc$ and $R_{\rm{nb,2}}=220\pc$, 
to fit the data. The authors claimed that the presence of two distinct
components may not have a clear physical meaning, but could 
reflect the extinction effects not accounted for
in their model. By assuming $\Upsilon=2$, \citet{launha_etal_02} derived the total mass of 
the nuclear bulge to be $(1.4\pm0.6)\times10^9\Msun$.

Following \citet{launha_etal_02}, the nuclear bulge 
included in our simulations is also composed of 
two components with $R_{\rm{nb,1}}=120\pc$ and $R_{\rm{nb,2}}=220\pc$. 
Their density profiles are described by Eq.~\ref{eq:nbprofile}. 
We generally assume the mass-to-light ratio $\Upsilon$ is the 
same for the two components. The total mass of the nuclear bulge 
is therefore $1.4\times10^9\Msun$ with $\Upsilon_1=\Upsilon_2=2.0$, 
consistent with \citet{launha_etal_02}. In Section~\ref{sec:nbresults} 
we test a slightly different nuclear bulge mass model 
where $\Upsilon_1=3.46$ and $\Upsilon_2=0$.

As the nuclear bulge is a massive and compact object, its dynamical effects cannot be negligible. 
Other studies have also revealed the presence of the nuclear bulge 
(or at least a massive and dense central object), including
the young stellar nuclear disk found in APOGEE by \citet{schonr_etal_15}, 
and the increase of stellar velocity dispersion caused by a massive 
central object in \citet{portail_etal_17} and \citet{martin_etal_18}. 
\citet{li_etal_16} also included the nuclear bulge to 
generate a reasonable CMZ in their best-fit gas dynamical models. 
However, the exact mass of the nuclear bulge is still not well determined. 
This is partly because that the 
COBE photometry used to derive the density profile of the nuclear bulge  
is heavily affected by the dust extinction. Another reason is that 
the assumed $\Upsilon=2$ in the near-infrared band 
is uncomfortably large \citep[e.g.][]{conroy_etal_09,con_gun_10}. 
We therefore keep $\Upsilon$ as a free parameter in this study.

\subsubsection{Soliton core}
\label{sec:pot_sc}

\begin{figure}[!t]
\includegraphics[width=0.5\textwidth]{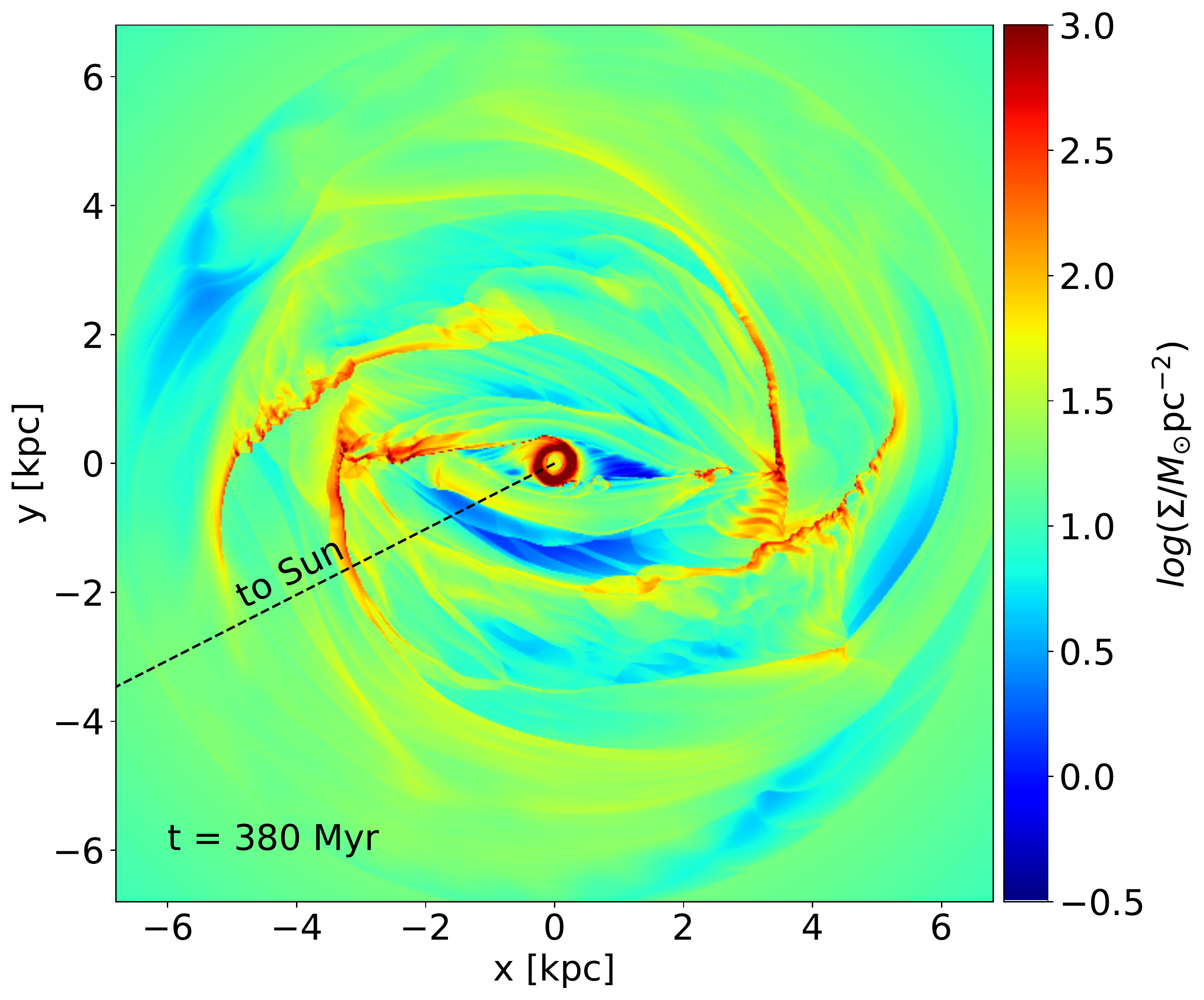}
\caption{The gas surface density $\Sigma$ 
in our hydrodynamical simulations with $\cs = 10\kms$. 
The bar lies horizontally along $x$-axis with a semi-major 
length around $\approx5\kpc$, and 
the corotation radius of the bar is placed 
at $R=6.2\kpc$. 
The CMZ can be seen as the high density gas ring around $R\approx300\pc$.
The central potential in this model 
includes a nuclear bulge with $\Upsilon=2$ but without a soliton core.  
The Sun is at $(x,y)=(-7.4\kpc,-3.8\kpc)$ in this plot. The Sun-Galactic center 
line ($l=0$) is indicated by the black dashed line. 
\label{fig:overallgas}}
\vspace{0.2cm}
\end{figure}

FDM simulations found a tight relation between the central soliton 
core properties and the host halo mass \citep[]{schive_etal_14b}, although 
this relation may not be very accurate to predict the soliton core properties 
in the Milky Way, as noted in the introduction. 
The virial mass of the Milky Way halo has been constrained to 
be around $0.9-2.0\times10^{12}\Msun$ \citep[e.g.][]{portail_etal_17,pos_hel_19}.
Applying the halo-core relation in \citet{schive_etal_14b} with $m_{22}=1$ 
and $M_{\rm halo}=1.5\times10^{12}\Msun$, 
the expected soliton mass $M_s$ and core radius $r_{c}$ in the Milky Way are 
$M_s = 1.44\times10^9\Msun$ and $r_c=0.16\kpc$, respectively. 
Here $m_{22}$ is the boson particle mass in unit of $1.0\times10^{-22}\eV$, and 
$r_{c}$ is the core radius defined as the radius enclosing $25\%$ 
of the total soliton mass $M_s$ ($r_c$ is also roughly the 
radius where density decreases to half of the central peak value).
However, as pointed out in Section II.C from
\citet{hui_etal_17}, these two numbers should be 
interpreted only as the upper and lower limits, 
i.e. the maximum soliton mass and the minimum soliton 
core radius since the Milky 
Way halo may have not been fully relaxed and 
the soliton core may be developing. 
We therefore define $M_{\rm smax} = 1.44\times10^9\Msun$ 
as the maximum soliton mass inside the Milky Way by 
adopting $m_{22} = 1$. 
Note that this mass is very close 
to the mass of the nuclear bulge with $\Upsilon=2$.

The mass $M_s$ and core radius $r_c$ of self-gravitating soliton cores 
are inversely related, since the characteristic de Broglie wavelength
cannot exceed the virial radius of the system \citep[][also see 
Eq.~\ref{eq:rcdef} in \S\ref{sec:highboson}]{hui_etal_17}. 
This relation needs to be modified for the Milky Way, due to 
the presence of the nuclear bulge whose mass and size are similar to 
the expected soliton core predicted by the halo-core relation. 
Recently, \citet{bar_etal_19} studied how the baryonic nuclear 
bulge would affect the soliton core size in the Milky Way. 
They found that the soliton core radius $r_c$ could be 
compressed to be half of its original value 
in the potential of a nuclear bulge with a mass profile 
similar to \citet{launha_etal_02}. 
Inspired by their work, 
we adopt a modified soliton core density profile based 
on the results in \citet{schive_etal_14b} as:
\begin{equation}
\rho_{\rm{soliton}}(r) = \frac{0.082~(r_c/{\rm kpc})^{-3}(M_s/10^9\Msun)}
                         {[1+9.1\times10^{-2}~(r/r_c)^2]^8}~\Msun\pc^{-3},  
\label{eq:scprofile}
\end{equation}
where $M_s$ and $r_c$ are \emph{independent}. We could then use Eq.~\ref{eq:scprofile} 
to make soliton cores with the same mass but different core radius. 
The corresponding gravitational acceleration of such a soliton core 
is given by: 
\begin{multline}
a_{\rm{soliton}}(r) = 1.82\times10^{-4}~G~M_s \frac{1}{r^2(b^{2} + 1)^{7}}  \\
                      (3465 b^{13} + 23100 b^{11} + 65373 b^{9} + 101376 b^{7} + 92323 b^{5} \\
                      + 48580 b^{3} - 3465 b + 3465 (b^{2} + 1)^{7} \arctan{(b)}),
\label{eq:scrotcurve}
\end{multline}
where $b=(2^{1/8}-1)^{1/2} (r/r_c)$ and $G=4.302\kpc~(\kms)^2~(10^6\Msun)^{-1}$.

Assuming $M_s$ and $r_c$ are independent 
implies the mass profile of the nuclear bulge varies 
accordingly. This parametrisation is more flexible 
to obtain the required mass profile of the soliton core, 
while the method in \citet{bar_etal_19} is a more 
self-consistent way to determine the soliton core properties 
by solving the Schr{\"o}dinger-Poisson equation. 
In this study, we first adopt $M_s$ and $r_c$ 
as two free soliton parameters. We then  
examine whether the preferred soliton core 
properties (i.e. certain combinations of $M_s$ and $r_c$) 
are allowed by the nuclear bulge potential we adopt in \S\ref{sec:highboson}. 

If $M_s$ and $r_c$ are chosen to be the same as the predictions 
of the halo-core relation \citep{schive_etal_14b}, one gets a self-gravitating
soliton core. If $M_s$ is the same but a smaller $r_c$ is adopted, 
one gets a compressed soliton core under an external baryonic potential 
(i.e. the nuclear bulge). In Fig.~\ref{fig:scrc} we show the 
rotation curves obtained by Eq.~\ref{eq:scrotcurve} of a 
self-gravitating soliton core and 
a compressed soliton core with the same total mass.  
Our result is very similar to Fig. 4 in \citet{bar_etal_19}. 
Note that $M_s=M_{\rm smax}=1.44\times10^9\Msun$ 
and $r_c=0.16\kpc$ correspond to the original halo-core 
relation predictions for a $M_{\rm halo}=1.5\times10^{12}\Msun$ 
Milky Way mass halo and $m_{22}=1$ \citep{schive_etal_14b}.

\begin{figure*}[!t]
\includegraphics[width=1.05\textwidth]{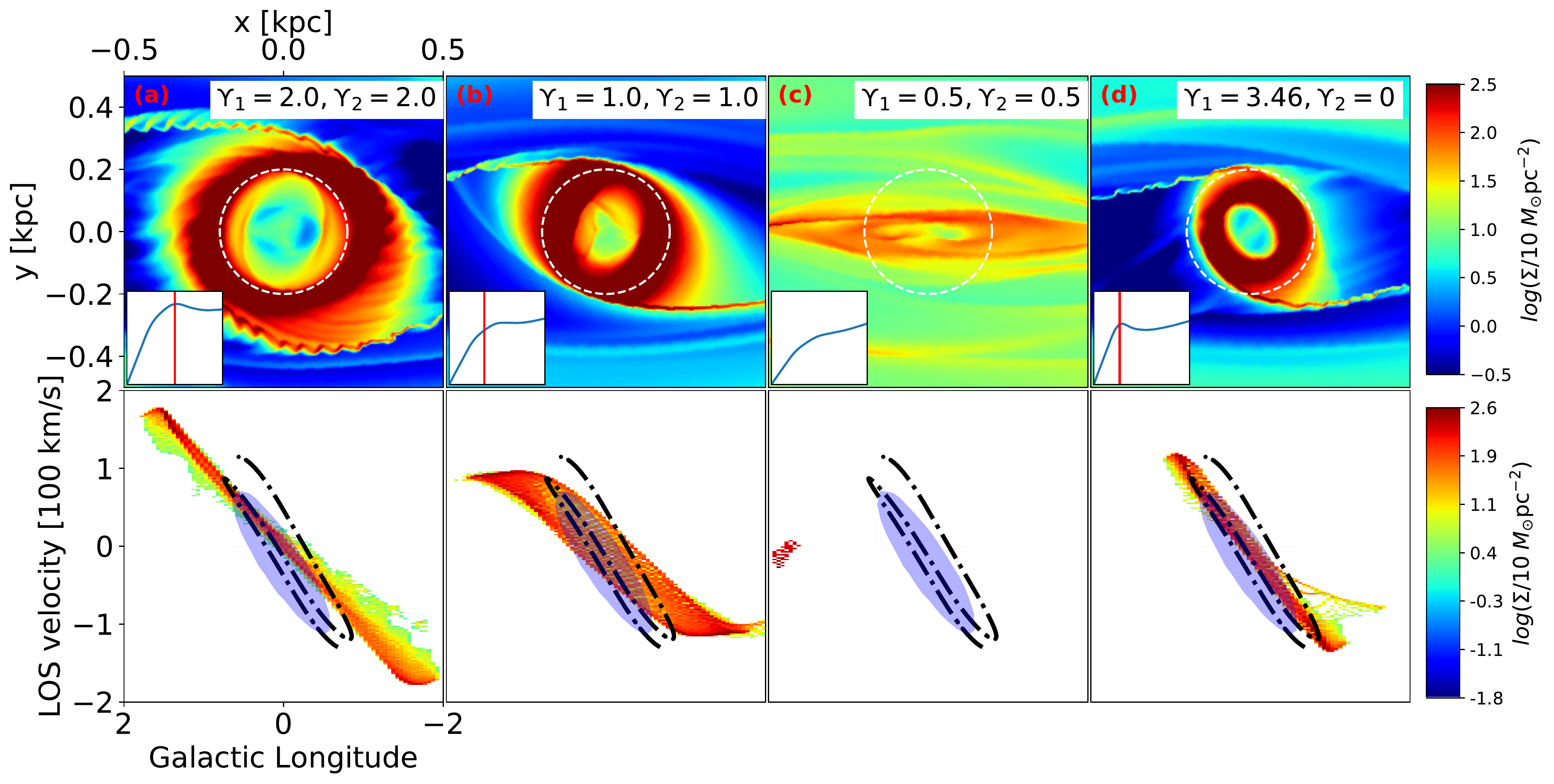}
\caption{Effects of the mass-to-light ratio $\Upsilon$ of the nuclear bugle 
on the shape and kinematics of the CMZ, 
with a sound speed $\cs=10\kms$. 
No soliton cores have been included in these four models. 
The snapshots are taken at $T=500\Myr$.
Top panels: gas surface density 
in the inner $500\pc$. From left to right $\Upsilon$ is 
decreased from $2$ to $0.5$, corresponding to a mass from $1.4\times10^9\Msun$ 
to $3.5\times10^8\Msun$. The white dashed circle denotes a radius of $200\pc$,
representing the outer boundary of the CMZ.
The inset 
at the bottom left corner shows 
the rotation curve (blue line) and the 
averaged ring radius (vertical red line) in the model. 
The boundary of the inset 
is $500\pc$ for $x$-axis and $200\kms$ for $y$-axis. 
The inset in panel (c) does 
not show a ring radius since the density of gas 
is not high enough to be called a CMZ (criteria: 
$\Sigma \ge 2\times 10^3\Msun\pc^{-2}$). 
Bottom: \lvplots\ of the models, which show 
the distribution of gas density as a function 
of Galactic longitude $l$ and line-of-sight velocity $v$.
The limits of the colorbar are chosen to 
highlight most of the features in $(l,v)$ space. 
The black dotted-dashed 
line is the open stream model of the CMZ proposed 
by \citet{kruijs_etal_15}. The 
blue shaded region shows the [${\rm C}_{\rm \RNum{2}}$]
observations in CMZ from \citet{langer_etal_17}.
The bottom part of panel (c) is almost empty because 
the there are few dense gas in this model 
that satisfies our CMZ criteria.  
Note that Galactic longitude of $2^\circ$ corresponds to a 
radius of $\approx300\pc$. 
\label{fig:nbcmz}}
\vspace{0.2cm}
\end{figure*}

\subsection{Numerical Scheme}
\label{sec:simulationdetial}

We study how a thin gas 
disk evolves under a realistic barred Milky Way potential 
described in the previous sections. The barred potential is rigidly rotating about 
the Galactic center with a fixed pattern
speed $\Omega_b = 33\freq$. 
The gas is assumed to be in-viscid, and we use the latest public version of 
the grid-based MHD code \textit{Athena++} (Stone et al. in preparation) 
to solve the Euler equations. \textit{Athena++} is a complete re-write 
of the previous \textit{Athena} code 
with improved performance and scalability. 
We employ a uniform Cartesian grid with $4096\times4096$ cells for 2D simulations
covering a square box with size of $L=14\kpc$, thus the
grid spacing is $\Delta x=\Delta y=3.4\pc$. We also test 3D runs where the grid is 
$2048\times2048\times21$ with a vertical extent of $200\pc$. We adopt the
\textit{roe} Riemann solver, piece-wise linear reconstructions, and outflow boundary
conditions at the domain boundaries. High spatial resolution is necessary to 
capture the turbulent gas motions in the central region of galaxies, as demonstrated in previous works
\citep[e.g.][]{sorman_etal_15a, li_etal_15, few_etal_16}. 
All our models are run for a period of $0.5\Gyr$. The 2D simulations take about
$\sim15-20$ hours using 64 INTEL cores, and the 3D runs take about 
$\sim5$ days using 256 INTEL cores. In general, \textit{Athena++} is $\sim5$ times
faster than the previous version \textit{Athena} \citep{gar_sto_05, stone_etal_08, sto_gar_09}. 

We set up the initial rotating gaseous disk with an 
exponential surface density profile
$\Sigma_{\rm {gas}} = \Sigma_0\exp{(-R/R_{\rm {d}})}$, 
where $\Sigma_0=76.7\Surf$ and $R_{\rm {d}} = 4.8\kpc$. 
The gas disk then has a surface density around $13\Surf$ at
solar neighbourhood to match the observed value roughly \citep[]{bov_rix_13}. 
The initial rotation velocity 
of gas is set to balance the azimuthally averaged gravitational 
force. We linearly ramp up the 
bisymmetric bar potential over the first bar rotation period $T_{\rm {bar}}=186\Myr$ 
to avoid transients. We compute the gas
flow in the inertial frame, while \citet{li_etal_16} computed it in the
bar corotation frame. 

We adopt an isothermal equation of state assuming the specific gas internal energy is 
constant by an energy balance
between radiative heating and cooling processes, same as previous 
studies \citep[e.g.][]{kim_etal_12a,kim_etal_12b,sorman_etal_15c}. 
The effective isothermal sound speed $\cs$ describes the level of 
velocity dispersion between molecular clouds, thus does not 
stand for the microscopic gas temperature. 
We test two different $\cs$ in our models 
with low ($10\kms$) and high ($20\kms$) cases, in the range of  
the observed ISM velocity dispersion 
\citep[e.g.][]{walter_etal_08,leroy_etal_09}. 
We neglect gas self-gravity, magnetic fields, 
star formation and stellar feedback, 
and other additional physics in our models for simplicity, 
but prefer to study these effects in a follow-up paper.
We discuss how these unaccounted physics will affect our results in 
\S\ref{sec:otherphysics}.

\begin{figure*}[!t]
\includegraphics[width=1.0\textwidth]{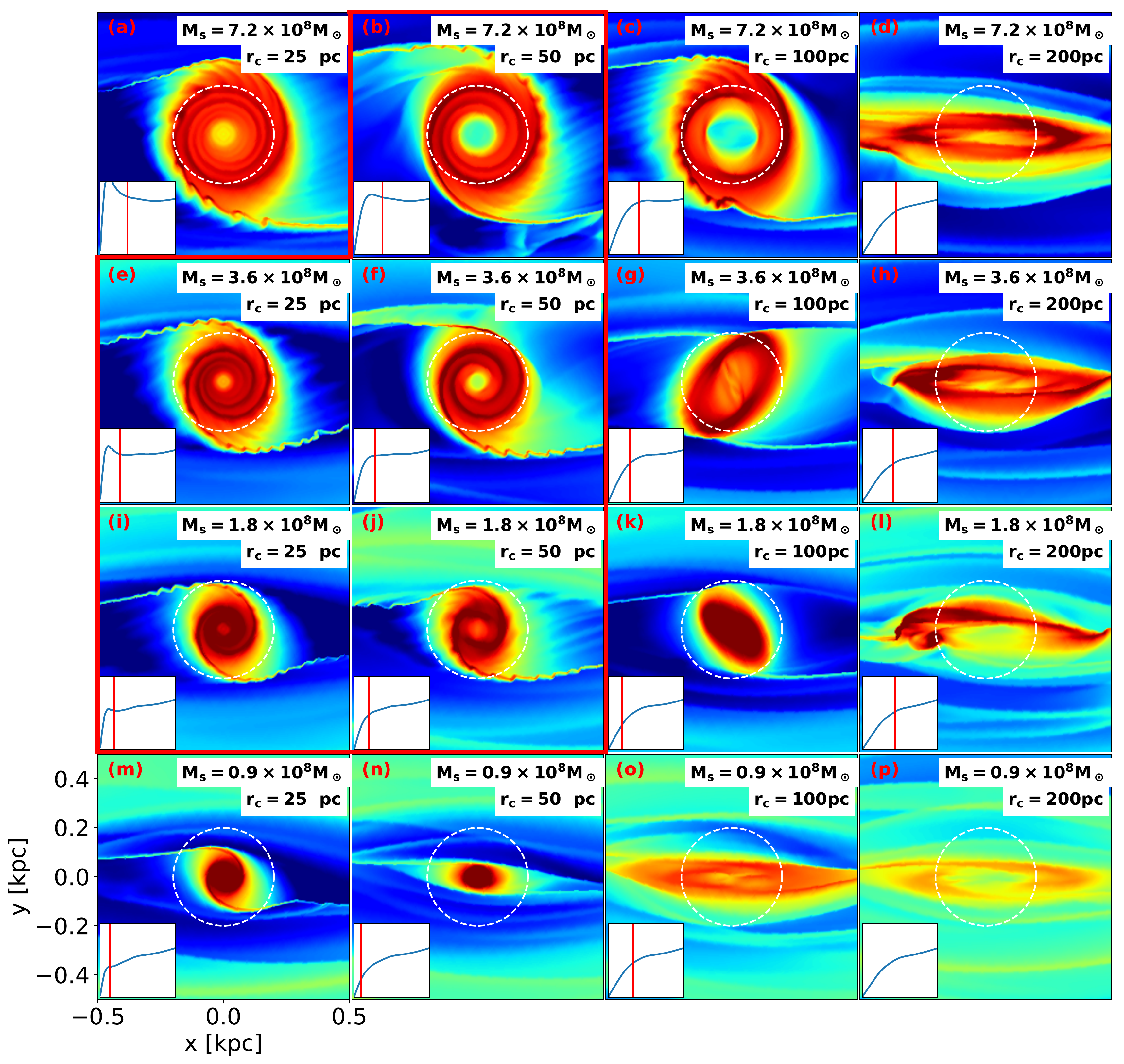}
\caption{
Gas surface density in the inner $500\pc$ for 
the models that include a nuclear bulge with $\Upsilon=0.5$ 
and a soliton core with different mass and size. 
The lines, symbols, and the color map 
are identical to those in Fig.~\ref{fig:nbcmz}. 
The snapshots are taken at $T=500\Myr$.
From left to right the soliton 
core radius $r_c$ increases from $25\pc$ 
to $200\pc$, and from top to bottom the mass of the 
soliton core $M_s$ 
decreases from $7.2\times10^8\Msun$ ($1/2M_{\rm smax}$) 
to $0.9\times10^8\Msun$ ($1/16M_{\rm smax}$). 
$M_{\rm smax} = 1.44\times10^9\Msun$ 
is the maximum soliton mass inside a 
$1.5\times10^{12}\Msun$ Milky Way halo predicted 
by the halo-core relation in \citet{schive_etal_14b} assuming $m_{22}=1$. 
Note that $M_{\rm smax}$ is very close to the mass of 
the nuclear bulge derived in \citet{launha_etal_02}. 
The sound speed $\cs$ is $10\kms$. The inset in the bottom 
right panel does not have a ring radius since the 
density of gas is not high enough, similar to Fig.~\ref{fig:nbcmz}. 
The red box highlights the models that roughly match the observed 
CMZ size. 
\label{fig:sccmzden}}
\vspace{0.2cm}
\end{figure*}

\begin{figure*}[!t]
\includegraphics[width=1.0\textwidth]{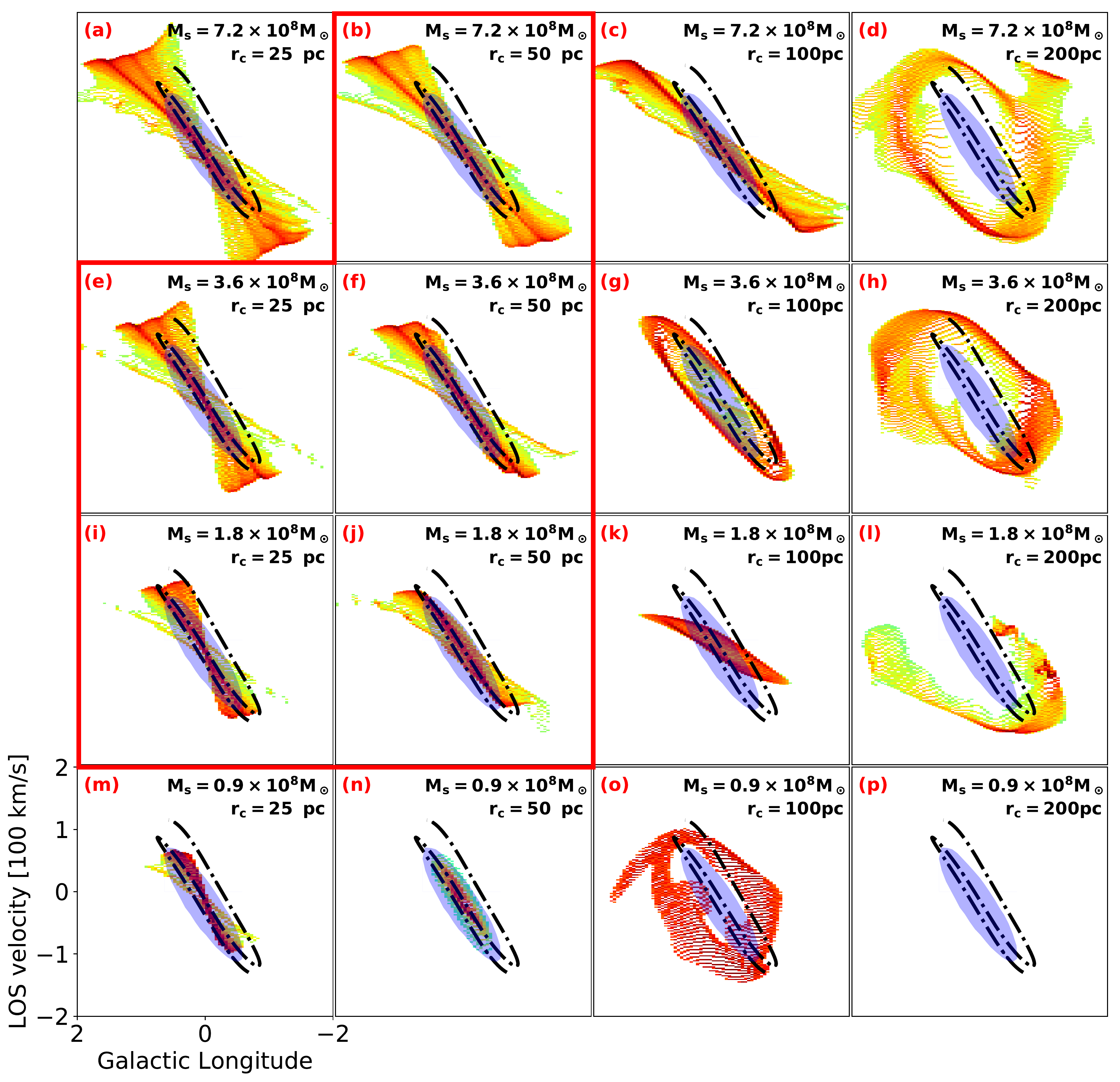}
\caption{The \lvplot\ in the inner $2^\circ$ of Galactic longitude $l$ 
for the models shown in Fig.~\ref{fig:sccmzden}.
The lines, symbols, and the color map are identical to those in Fig.~\ref{fig:nbcmz}.
The bottom right panel is almost empty because 
the density of gas is not high enough in this model, similar to Fig.~\ref{fig:nbcmz}. 
The red box highlights the models that can roughly reproduce the observed 
CMZ kinematics. 
\label{fig:sccmzlv}}
\vspace{0.2cm}
\end{figure*}

\section{Simulation results}
\label{sec:simulationresults}

\subsection{General evolution}

We first report the simulation of the gas evolution in the potential
described in \S\ref{sec:pot_general} with a $\Upsilon=2$
nuclear bulge, but without a soliton core. We explore the 
case with a relatively low sound speed $\cs=10\kms$, which is 
the same as most of the previous studies 
\citep[e.g.][]{fux_99b,rod_com_08,ridley_etal_17}. In Fig.~\ref{fig:overallgas} 
we show the gas surface density in the simulation at $t=380\Myr$ when 
the bar has finished about two rotation period. 
Strong shocks develop at the leading side of the bar as the flow is supersonic, 
which are indicated by the high density gas ridges in Fig.~\ref{fig:overallgas}. 
Gas loses energy and angular momentum when encountering the shocks, 
then flows inwards and accumulates at center. The CMZ can be clearly seen 
as the high density ring-like structure at $R\approx300\pc$ in the center. 
The four spiral arms around the bar region are known 
as the near and far 3-kpc arms and the molecular ring in the \lvplot,
which are related to the 4:1 resonance of the rotating bar \citep{sorman_etal_15b}. 
The flow pattern then becomes quasi-steady until the end of
the simulation.

In general, this model is very similar to the best-fit model of 
\citet{li_etal_16}, as we adopt the same Galactic potential but 
with a higher resolution. In the following sections we vary the 
mass-to-light ratio $\Upsilon$ of the nuclear bulge, the mass of the 
soliton core $M_s$, and the radius of the soliton core $r_c$ 
to determine the possible mass and size of the soliton core 
allowed in the Milky Way, thus putting constraints on FDM theory. 

\begin{figure*}[!t]
\includegraphics[width=1.0\textwidth]{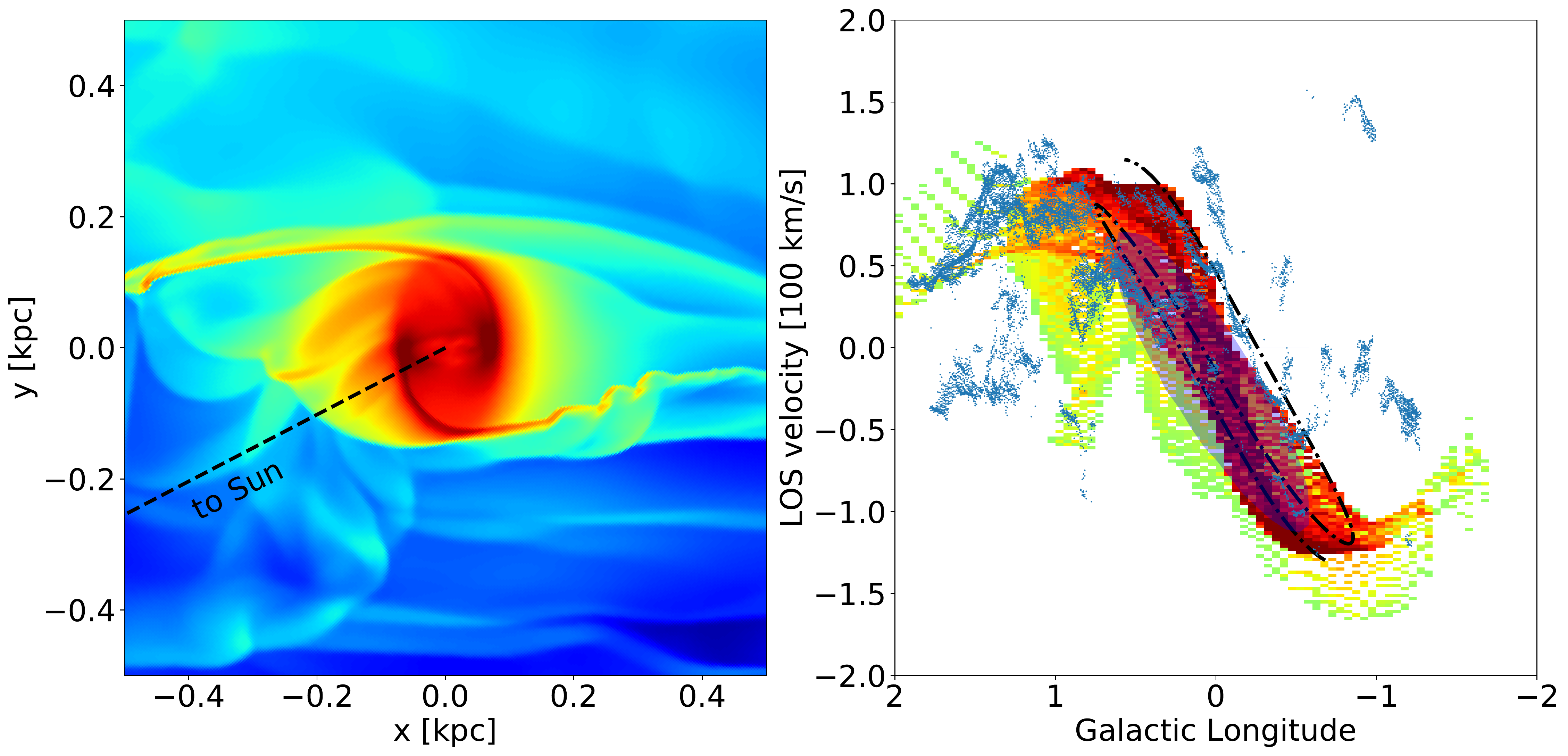}
\caption{Gas surface density and the corresponding \lvplot\ with $\cs=20\kms$.
The lines, symbols, and the color map are identical to those in Fig.~\ref{fig:nbcmz}. 
The model includes a nuclear bulge with $\Upsilon=0.5$ and a soliton core of  
$M_s=3.6\times10^8\Msun$ and $r_c=50\pc$. This model is 
almost identical to panel (f) in Fig.~\ref{fig:sccmzden} and \ref{fig:sccmzlv}, 
expect that the sound speed is twice higher. The blue dots in the right panel 
are NH$_3$ data from HOPS survey \citep{longmo_etal_17} fitted by the 
SCOUSE package \citep{hensha_etal_16}. The snapshot is taken at 
$T=500\Myr$. The \lvplot\ of this model 
resembles most of the observed features. 
\label{fig:sccmzcs}}
\vspace{0.2cm}
\end{figure*}

\subsection{Tests with only the nuclear bulge}
\label{sec:nbresults}

As we have shown in \S\ref{sec:pot_nb}, An assumed a mass-to-light ratio 
$\Upsilon = 2$ gives a nuclear bulge mass of $1.4\times10^9\Msun$, 
which is very close to the maximum mass of the soliton core 
$M_{\rm smax} = 1.44\times10^9\Msun$ with $m_{22}=1$. 
However, the total mass of the central 
components is constrained to be $\approx1.5\times10^9\Msun$ 
\citep[e.g.][]{launha_etal_02,portail_etal_17,martin_etal_18}. 
It is therefore not reasonable 
that the Milky Way hosts both a maximum mass soliton core and a 
$\Upsilon = 2$ nuclear bulge at the same time. 
The mass degeneracy of a soliton core and a stellar nuclear bulge 
in the Milky Way center was also discussed in \citet{bar_etal_18}. 
As $\Upsilon$ of the nuclear bulge may 
have some uncertainties, we first examine its effects on gas 
features when a possible soliton core is not included. 

In Fig.~\ref{fig:nbcmz} we plot the gas surface density (top panels), 
the rotation curves in the central $500\pc$ with the position of the CMZ marked (insets), 
and the corresponding \lvplot\ (bottom panels) of the gas in the central 
$500\pc$ of our simulations with different $\Upsilon$. 
The colors in the \lvplot\ indicate gas density which is binned according to 
Galactic longitude and line-of-sight velocity. 
We only select the dense 
gas ($\Sigma \ge 2\times 10^3\Msun\pc^{-2}$) 
to represent the CMZ and show their 
corresponding $(l,v)$ features, therefore 
the colors in \lvplots\ are mostly red. 
The blue shaded region in the bottom panels 
indicates [${\rm C}_{\rm \RNum{2}}$] observations 
in the CMZ region from \citet{langer_etal_17}, while the dotted-dashed line 
is the open stream model of the CMZ obtained by 
fitting the NH$_3$ data \citep{kruijs_etal_15}. 
The [${\rm C}_{\rm \RNum{2}}$] mainly traces the hot ionized gas, 
while NH$_3$ originates from cold and dense molecular gas. It seems that 
gas with different physical conditions in the CMZ 
forms a consistent shape in the $(l,v)$ space. 

To form a CMZ that has similar size and kinematics compared to observations, 
the model needs to satisfy (at least) two conditions: 
(1) the rotation curve in the model has a circular velocity of 
$\approx100\kms$ at $R\approx100\pc$; (2) the inflowed gas piles up on roughly 
circular orbits with $R\approx100\pc$. In Fig.~\ref{fig:nbcmz} we show 
that the size of the CMZ shrinks with lower $\Upsilon$, due to a decreasing central mass 
\citep[e.g.][]{athana_92b,reg_teu_03,li_etal_15,li_etal_17}. For $\Upsilon = 0.5$, 
a central mass of $3.6\times10^8\Msun$ is not sufficiently massive to
produce a nearly-circular CMZ with high gas density. 
The model with $\Upsilon = 2$ roughly satisfies condition (1), but the radius of 
CMZ is larger than the preferred value. The white dashed circle in the plot has a radius 
of $200\pc$, which indicates the outer boundary of the CMZ \citep[e.g.][]{molina_etal_11,sorman_etal_19}. 
On the other hand, the model with $\Upsilon = 1$ 
roughly satisfies condition (2), but the rotation velocity is lower, 
and thus it forms a shallower slope in the \lvplot\ compared to observations. 
These results imply that varying $\Upsilon$ of the nuclear bulge 
alone with the density profile given by Eq.~\ref{eq:nbprofile} may 
not produce a reasonable CMZ that is consistent with observations 
in both size and kinematics. 

We slightly fine-tune the model of the nuclear bulge above
to see whether we could better reproduce the observed 
CMZ kinematics. We find a less massive but more concentrated 
nuclear bugle is needed to satisfy conditions (1) and (2) 
simultaneously. Fig.~\ref{fig:nbcmz}(d) shows such a model 
with $\Upsilon_1=3.46$ for the first component ($R_{\rm{nb,1}}=120\pc$), 
and $\Upsilon_2=0$ for the second component ($R_{\rm{nb,2}}=220\pc$). 
This results in a nuclear bulge with a mass of
$5.1\times10^{8}\Msun$. The gas kinematics in Fig.~\ref{fig:nbcmz}(d) agree with observations 
fairly well. We regard this model which does not include a soliton core
as one of our best-fit models, although the nuclear bulge density profile in this 
model is slightly different from that in \citet{launha_etal_02}. 
If the luminosity profile in \citet{launha_etal_02} is correct, then this 
best-fit model implies that the mass-to-light ratio of the nuclear bulge 
is radially varying.  

\subsection{Tests with possible soliton cores}

In \S\ref{sec:pot_sc} we argue that the soliton core 
(if it exists) could be significantly compressed by the 
presence of a nuclear bulge. The compressed 
soliton core would result in a steeper rotation curve at the inner 
region, which helps to generate a steeper slope in the $(l,v)$ space for 
the CMZ. In the following sections we adopt the nuclear bulge profile 
as determined in \citet{launha_etal_02}, where the mass-to-light 
ratio is the same for the two components. As shown in \S\ref{sec:nbresults}, 
such a nuclear bulge is not concentrated enough, thus leaves some room for
a compressed soliton core. 

Fig.~\ref{fig:nbcmz}(b) suggests a $\Upsilon = 1$ 
nuclear bulge already generates a slightly larger CMZ compared 
to observations. Adding a compact soliton core together with such a nuclear
bulge would only make the CMZ even larger. We thus adopt $\Upsilon = 0.5$ 
for the nuclear bulge so that we can produce a CMZ smaller than 
$200\pc$ by including a suitable soliton core. 


Fig.~\ref{fig:sccmzden} and 
Fig.~\ref{fig:sccmzlv} present the gas surface density and the 
corresponding \lvplot\ for the models with different soliton core mass $M_s$
and core radius $r_c$, together with a $\Upsilon = 0.5$ nuclear bulge. 
From left to right the core radius $r_c$ are $25\pc$, $50\kpc$, $100\pc$
and $200\pc$, respectively. From top to bottom the soliton core mass 
$M_s$ are $7.2\times10^8\Msun$ ($1/2M_{\rm smax}$), 
$3.6\times10^8\Msun$ ($1/4M_{\rm smax}$), $1.8\times10^8\Msun$ 
($1/8M_{\rm smax}$), and $0.9\times10^8\Msun$ ($1/16M_{\rm smax}$). 
$M_{\rm smax} = 1.44\times10^9\Msun$ 
is the maximum soliton mass inside a 
$1.5\times10^{12}\Msun$ Milky Way halo predicted 
by the halo-core relation in \citet{schive_etal_14b} assuming $m_{22}=1$.
The blue line at the inset of each panel 
denotes the rotation curve, and the vertical red line is the averaged CMZ radius, 
same as in Fig.~\ref{fig:nbcmz}. 

It is clear that the location of CMZ is roughly the the turnover 
radius of the rotation curve (i.e. where the rotation curve turns flat). 
For the detailed formation mechanism we refer the reader to 
\citet{li_etal_15}. The effects of different soliton core mass $M_s$ 
and core radius $r_c$ 
are similar to those of varying $\Upsilon$ of 
the nuclear bulge. A larger $M_s$ leads to a larger CMZ, while 
a smaller $r_c$ produces a more circular CMZ together 
with a steeper $(l,v)$ feature. Combing Figs.~\ref{fig:sccmzden} and 
\ref{fig:sccmzlv}, we conclude that 
cases (b), (e), (f), (i) and (j) 
can roughly satisfy conditions (1) and (2) mentioned in \S\ref{sec:nbresults} 
simultaneously. The CMZ formed in these five simulations better reproduce 
the observed gas kinematics together with a size of $\approx200\pc$.
These five models are highlighted with a red box in the figures, and 
are regarded as our best-fit models that contain a soliton core. 

The soliton cores with larger $r_c$ form more elliptical CMZs compared 
to models with smaller $r_c$. The orientation of these more elliptical CMZs 
oscillates with time, which is not shown here but can be inferred in 
panel (g) and (k) in Fig.~\ref{fig:sccmzden}. This is because that gas tend not to 
follow periodic orbits when the pressure force is comparable to the gravitational 
force. The oscillating CMZ is more or 
less similar to the open stream model proposed by \citet{kruijs_etal_15}, 
who argued that a processing orbit instead of a closed orbit is 
a more physical solution in a Galactic potential. In our models, when 
these elliptical CMZs have a specific angle with respect to the Sun, 
the resulting $(l,v)$ plot agrees better with observations (e.g. 
panel g), while other angles are worse (e.g. panel k). 
This can be understood by the following arguments: 
The Sun is at $(x,y)=(-7.4\kpc,-3.8\kpc)$ in our models (also see 
the dashed Sun-Galactic center line in Fig.~\ref{fig:sccmzcs}), 
if the line-of-slight is tangent to the elliptical CMZ near its pericenter  
(e.g. panel g), we would expect a steeper slope in 
the $(l,v)$ space due to a higher LOS velocity and a narrower extent 
compared to the case where the tangent point is near the apocenter 
(e.g. panel k). It is possible that the shape of the 
CMZ changes with time in reality, but it then implies 
we are observing the CMZ at a special time. If this is true, 
the constraints on the soliton core radius $r_c$ 
could be loosened to a slightly larger value. 

\subsection{Varying sound speed}

Apart from the gravitational constraints, 
the gas in the CMZ is also quite turbulent. ALMA observations of 
SiO revealed the velocity dispersion of gas can be as large as 
$\sim20\kms$ \citep{kauffm_16}. The NH$_3$ data from HOPS 
survey \citep{longmo_etal_17} indicate a velocity 
dispersion of $17.4 \pm 4.8\kms$. As described in 
\S\ref{sec:simulationdetial}, the effective 
isothermal sound speed $\cs$ used in our simulations 
reflects mainly the velocity dispersion of gas clouds. We therefore further test 
a higher value $\cs=20\kms$ for the preferred models highlighted by 
the red box in Fig.~\ref{fig:sccmzlv} to see how this would affect our results.  

We find that a higher sound speed $\cs$ generally makes the CMZ smaller, but 
only slightly. This is because shocks with a higher $\cs$ are developed 
closer to the bar major axis. The physical reason is that the gas is 
less supersonic thus needs a steeper potential to be shocked 
\citep[e.g.][]{kim_etal_12a,sorman_etal_15a}. 
We also find that the CMZ tends to be more elliptical in a higher $\cs$ case, similar 
to the model with a large core radius $r_c$ but a lower $\cs$. A higher 
$\cs$ also helps to broaden the $(l,v)$ features as a result of a higher 
velocity dispersion, which better matches the data. 
Fig.~\ref{fig:sccmzcs} shows the model that matches most observed $(l,v)$ features. 
The model hosts a nuclear bulge with $\Upsilon=0.5$, together with a soliton core of  
$M_s=3.6\times10^8\Msun$ and $r_c=50\pc$. This model is 
almost identical to panel (f) in Fig.~\ref{fig:sccmzden} and \ref{fig:sccmzlv}, except 
with a $\cs=20\kms$ instead of $10\kms$. The dashed line in the left 
panel indicates the Sun-Galactic center line (i.e. $l=0\degree$). 
The blue dots on the right panel show the NH$_3$ observations in the CMZ 
from HOPS survey \citep{longmo_etal_17} fitted by the 
SCOUSE package \citep{hensha_etal_16}. 

It is interesting that the model shown in Fig.~\ref{fig:sccmzcs} 
also qualitatively reproduces the asymmetry of the CMZ, 
i.e. it shows more emission on the positive longitude compared to the 
negative longitude in the $(l,v)$ space. From the face-on view, 
we can see that the gas forms some feather-like features on the left 
side of the CMZ. This may be caused by the 
``wiggling'' instability proposed by \citet{wad_kod_04} then further studied 
by \citet{kim_etal_14} and \citet{sorman_etal_17b}. 
A higher $\cs$ enhances the instability, thus 
making the asymmetry more prominent than the lower $\cs$ case. Some of the 
``feathers'' roughly reproduces the observed Sagittarius B2 cloud at $l\approx1\degree$. 
It is worth noting that the asymmetry is transient, with a typical timescale of a 
few tens Myr, similar to some of the more sophisticated simulations 
\citep[e.g.][]{sorman_etal_17a}.

We thus conclude that a model with a light and compact 
soliton core (i.e. $M_s\approx3.6\times10^8\Msun$ 
and $r_c\approx50\pc$), together with a $\Upsilon=0.5$ 
nuclear bulge can roughly reproduce 
most of the observed CMZ properties. The match is better when 
adopting a higher sound speed $\cs\approx20\kms$.
Nevertheless, it is still worth exploring for a better match to 
the observed features in the future studies.

\begin{figure}[!t]
\includegraphics[width=0.52\textwidth]{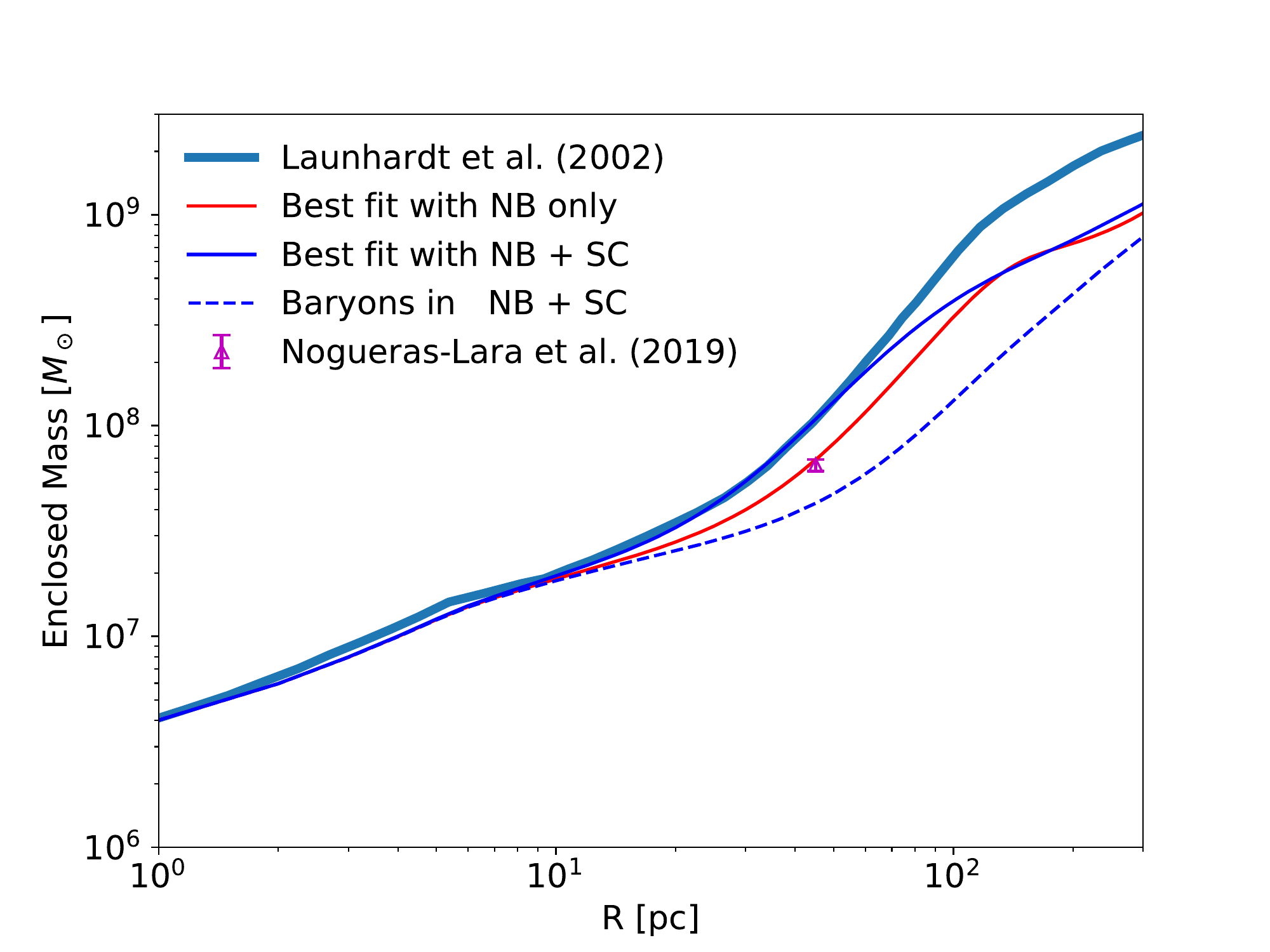}
\caption{Enclosed mass in spheres of radius $R$ for the inner
$300\pc$ of the Galaxy. The thick blue line is the result from 
Fig.14 in \citet{launha_etal_02}. The red solid line shows the 
mass distribution in Fig.~\ref{fig:nbcmz}(d), which is our 
best-fit model that includes only the nuclear bulge. The blue 
solid line shows that used in Figs.~\ref{fig:sccmzden}(f) and \ref{fig:sccmzlv}(f), 
which is one of our best-fit models that contains 
both the nuclear bulge and the soliton core. 
The blue dashed line plots the baryonic mass in this model. The purple 
triangle is the stellar mass ($6.5\pm0.4\times10^7\Msun$) 
obtained by \citet{noguer_etal_19} at $45\pc$. Note that this mass is not the enclosed 
mass inside a $r=45\pc$ sphere, but the total mass inside a 
$R=45\pc$ and $z=20\pc$ cylinder. We also include a $2.6\times10^6\Msun$ 
black hole and a $3.0\times10^7\Msun$ nuclear stellar cluster in the models 
to be consistent with \citet{launha_etal_02}.
\label{fig:MWmass}}
\vspace{0.2cm}
\end{figure}

\subsection{What does a light and compact soliton core imply?}
\label{sec:highboson}

We use $M_{\rm smax} = 1.44\times10^9\Msun$ as the maximum 
soliton mass inside the Milky Way, which is predicted by the 
halo-core relation in \citet{schive_etal_14b} \emph{assuming $m_{22}=1$}. 
However, the accepted models in Figs.~\ref{fig:sccmzden} and 
\ref{fig:sccmzlv} prefer much smaller soliton core mass $M_s$ 
compared to $M_{\rm smax}$. In this section we demonstrate 
that these light and compact soliton cores 
correspond to a boson mass range of $m_{22}\approx2-7$. 

\citet{schive_etal_14b} presented the relation 
between self-gravitating 
soliton core properties $(M_{\rm smax}, r_{c,0})$ 
and $m_{22}$ as:
\begin{align}
M_{\rm smax}  & = 1.44~{m_{22}}^{-1}\times10^9\Msun, \nonumber \\
r_{c,0}  &= 0.16~{m_{22}}^{-1}\kpc,
\label{eq:scuncomprofile}
\end{align}
assuming the Milky Way halo mass is $1.5\times10^{12}\Msun$.
We use $r_{c,0}$ to denote the core radius of a soliton without compression. 
Once $m_{22}$ is fixed, $M_{\rm smax}$ and $r_{c,0}$ are 
determined for a given halo mass. A larger $m_{22}$ corresponds 
to a less massive and more concentrated soliton core. However, due to the 
compression by the nuclear bulge, the real core radius $r_c$ should be 
smaller than the original $r_{c,0}$, as we have shown in \S\ref{sec:pot_sc}. 
We therefore define a dimensionless 
compression parameter $C = r_c/r_{c,0}$ to describe the level 
of compression. The compressed core radius $r_c$ can be written as:
\begin{equation}
r_c = 231~C~\frac{10^9\Msun}{M_s}~m_{22}^{-2}\pc,
\label{eq:rcdef}
\end{equation}
which is independent of the halo virial mass. Substituting 
$C = r_c/r_{c,0}$ in Eq.~\ref{eq:rcdef} recovers 
the original results in \citet{schive_etal_14b}. Here $C=1$ corresponds to 
a self-gravitating soliton core, while 
for $C < 1$ the soliton core is compressed by baryons. 
Eqs.~\ref{eq:scuncomprofile} and 
\ref{eq:rcdef} suggest that for a given soliton mass and core radius $(M_s, r_c)$,
there is a degeneracy between the boson mass $m_{22}$ and the compression parameter $C$. 
In other words, both $C$ and $m_{22}$ can be varied accordingly to give 
the desired $r_c$ for a certain $M_s$.

We first show that the compression parameter $C$ in our best-fit models 
cannot be achieved by the background $\Upsilon=0.5$ 
nuclear bulge if $m_{22}$ is around 1. 
The parameter set $(M_s, r_c)$ of our best-fit models 
in Figs.~\ref{fig:sccmzden} and \ref{fig:sccmzlv}, 
i.e. panels (b, e, f, i, j), can be interpreted as 
$C=0.16,0.04,0.08,0.04,0.02$ with $m_{22}=1$, as suggested by 
Eq.~\ref{eq:rcdef}. We numerically solve the Schr{\"o}dinger-Poisson equation 
based on the method in \citet{bar_etal_19} to self-consistently derive the soliton core 
radius $r_c$ in the $\Upsilon=0.5$ nuclear bulge potential. 
We find $r_c$ should be around $160\pc$, 
$184\pc$, and $196\pc$ for a soliton core with mass of $1/2$, $1/4$, 
and $1/8 M_{\rm smax}$, assuming $m_{22}=1$. The corresponding $C$ is therefore 
$0.50,0.29,0.29,0.15,0.15$ for panels (b, e, f, i, j), about 
$3-7$ times larger than the numbers quoted above. This implies $m_{22}=1$ cannot 
produce the preferred soliton cores in Figs.~\ref{fig:sccmzden} and \ref{fig:sccmzlv}. 

On the other hand, a higher boson mass $m_{22}$ naturally 
produces less massive and more concentrated 
soliton cores, as both soliton core 
mass $M_s$ and core radius $r_c$ are inversely proportional to $m_{22}$. 
By solving the Schr{\"o}dinger-Poisson equation, we find $m_{22}=2.4$ 
could produce the compressed soliton core with $M_s=7.2\times10^8\Msun$ 
and $r_c=50\pc$ in Fig.~\ref{fig:sccmzlv}(b), under the $\Upsilon=0.5$ nuclear 
bulge potential we adopt. Similarly, Fig.~\ref{fig:sccmzlv}(e) corresponds 
to $m_{22}=5.0$, Fig.~\ref{fig:sccmzlv}(f) corresponds to $m_{22}=3.3$, 
Fig.~\ref{fig:sccmzlv}(i) corresponds to $m_{22}=7.0$, 
and Fig.~\ref{fig:sccmzlv}(j) corresponds to $m_{22}=4.4$. 

Considering all the models highlighted by the red box 
(i.e. b, e, f, i, and j) in Figs.~\ref{fig:sccmzden} 
and \ref{fig:sccmzlv}, we conclude that a boson mass $m_{22}$ in the range of  
$2-7$ is more reasonable in the current configuration of 
the Galactic potential that contains both the soliton core and the nuclear bulge. 

We need to emphasize that our upper limit for $m_{22}$ comes from 
the assumption that a minimum mass for the soliton core must exist, otherwise 
the CMZ will not form. However, as shown in \S\ref{sec:nbresults}, a nuclear bulge 
only model could also form a reasonable CMZ consistent with observed kinematics, 
as long as we assume the nuclear bulge to be 
more concentrated than that in \citet{launha_etal_02}. 
Therefore the upper limit we get (i.e. $m_{22}\le\sim7$) is 
only a weak constraint -- it could be larger than 7 if the 
nuclear bulge in reality is more compact in mass 
compared to the model we adopt in the current paper. 
On the other hand, the lower limit should be 
a stronger constraint (i.e. $m_{22}\ge\sim2$), as we have 
already assumed a relatively less massive nuclear bulge 
compared to previous studies.

\section{Discussions}
\label{sec:discussions}

\subsection{The possibility of a higher boson mass}
As pointed out above, a $m_{22}$ range of $2-7$ is preferred in this work.
This boson mass range is marginally higher than the conventional 
value $m_{22} \approx 1$ \citep[e.g.][]{schive_etal_14a}, 
but it is generally consistent with many astrophysical constraints, 
for example, classical dwarf spheroidal galaxies \citep[$m_{22}=1-2$;][]{chen_etal_17}, 
ultra-faint dwarf galaxies \citep[$m_{22}=3.7-5.6$;][]{cal_spe_16}, 
ultra-diffuse galaxy Dragonfly 44 \citep[$m_{22} \approx 3$;][]{wasser_etal_19}, 
stellar streams in the Milky Way \citep[$m_{22} >1.5$;][]{amo_loe_18}, 
supermassive black holes \citep[$m_{22}>2.0$ or $m_{22}<0.63$;][]{dav_moc_19}.
However, our $m_{22}$ range is still smaller than the 
results constrained by the rotation curves of low surface 
brightness galaxies \citep[$m_{22}\ge10$;][]{bar_etal_18,bar_etal_19}.
The larger boson mass also helps alleviate the tension arising 
from the high-z luminosity function
\citep[e.g.][]{schive_etal_16, corasa_etal_17, schive_chiueh_18} and Lyman-$\alpha$ forest \cite[e.g.][]{irsic_etal_17, leong_etal_19}, 
where either a larger boson mass or an extreme axion model is typically 
required to produce a sufficient amount of small-scale structure. 
Self-consistent FDM hydrodynamical simulations will be essential for narrowing 
down the mass constraints further. 

\subsection{Central mass profile of the Milky Way}

The CMZ size and kinematics constrain the 
total mass profile in the central $\sim200\pc$, 
but it is still not quite clear whether the central mass is
contributed by a compact stellar 
nuclear bulge, or a less compact nuclear bulge together with a 
dark soliton core. This mass degeneracy between a nuclear bulge and a soliton core 
was first discussed in \citet{bar_etal_18}, and a better determination on the 
mass-to-light ratio of the nuclear bulge (i.e. stellar mass) 
would be helpful to break this degeneracy. 
In Fig.~\ref{fig:MWmass} we plot the enclosed 
mass profiles for two of our best-fit models with and without the soliton core. 
The red solid line in Fig.~\ref{fig:MWmass} shows the enclosed mass 
profile from our nuclear bulge only model (Fig.~\ref{fig:nbcmz}(d)), 
the central mass in this model are contributed entirely by baryons. The blue solid line 
shows the enclosed mass profile from the model that contains both a soliton core and 
the nuclear bulge (Figs.~\ref{fig:sccmzden}(f) and \ref{fig:sccmzlv}(f)). The mass 
of the baryonic component (i.e. the nuclear bulge) in this model is plotted 
with the blue dashed line. We could see that the difference between the baryonic mass of 
these two models is only about a few $10^7\Msun$ 
at $R\approx50\pc$. Recently, \citet{noguer_etal_19} 
obtained the stellar mass in the central $45\pc$ to be $6.5\pm0.4\times10^7\Msun$ 
using the \textit{GALACTICNUCLEUS} survey, their results are indicated by the purple triangles 
with tiny error-bars in this plot. We see that this stellar mass is quite consistent with our nuclear 
bulge only model without a soliton core. If their mass estimation is accurate, then 
there is little room left for the FDM soliton core in the Milky Way center. Nevertheless, 
we still need further observations to provide tighter constraints 
on the baryon mass profile to better clarify this problem.

\subsection{Other dynamical effects of the soliton core}

We find that a soliton core with a mass of 
$\approx3.6\times10^8\Msun$ and a nuclear 
bulge with $\Upsilon=0.5$ would generate a CMZ that matches the observations 
relatively well. The total mass of these two components is therefore 
$\approx0.7\times10^9\Msun$. However, 
the velocity dispersion revealed by \textit{Gaia} and VVV surveys at 
$|l|\la10\degree$ requires a dense center of $\approx 1.5\times10^9\Msun$as found by 
\citet{martin_etal_18} and \cite{portail_etal_17}, which is about 
2 times larger than the value preferred here. 
We emphasize that the mass obtained in this paper are constraint 
mainly by the CMZ region. The CMZ size and kinematics 
offers no strong constraints on the mass profile outside 
$R\ga200\pc$ ($|l|\ga1.5\degree$). Therefore this discrepancy 
could be explained by additional mass outside the CMZ radius which 
could still result a large velocity dispersion. 
Indeed, the central velocity dispersion 
of the stars drops inside $R\la200\pc$ \citep{martin_etal_18}, 
implying a toroidal mass distribution of a less massive center and 
a more massive periphery. 

Interestingly, the soliton core could form such a 
toroidal shape due to the baryonic potential as reported in \citet{bar_etal_19}. 
We emphasize that a toroidal soliton core is just one possible solution. 
More detailed modelling is still needed to predict the exact soliton 
core properties in the Milky Way. 

\subsection{Different physical condition of the gas inside CMZ}
\label{sec:otherphysics}

The gas in the CMZ is multi-phase \citep[e.g.][]{langer_etal_17}, 
but it seems that both the 
cold and hot gas form a consistent feature in the $(l,v)$ space for which we 
are trying to match using our isothermal simulations. The isothermal equation of state 
is a simple assumption, and it forms a slightly smaller CMZ compared to the simulations with a 
complicated chemistry cooling network, as shown in \citet{sorman_etal_17a}. 
The cooling mechanism also helps to break the high density gas 
in the CMZ into small clumps seen in the 
NH$_3$ data, while in our simulation the gas is more smoothly distributed in the CMZ. The 
cooling time scale for the ISM is much shorter than the dynamical time, thus we expect the 
kinematics of the CMZ would not be significantly affected by considering the chemical evolution. 
The mass of the gas in CMZ is around $3.0\times10^7\Msun$ \citep[][]{launha_etal_02}, 
which is also small compared 
to the mass of the soliton core and the nuclear bulge used in this paper. 

In general, including self-gravity and cooling would make the CMZ larger 
\citep[e.g.][]{kim_etal_12b,sorman_etal_17a}, while including magnetic field, 
star formation, and stellar feedback would make it smaller \citep[e.g.][]{van_cha_09,kim_sto_12}. 
These physics basically influence condition (2) mentioned in \S\ref{sec:nbresults}, 
i.e. where the inflowed gas piles up.
Condition (1) is still determined mainly by the gravitational potential. A recent work by 
\citet{armill_etal_19} studied the star formation in the CMZ with a detailed modelling of 
gas physics and the stellar evolution. In their models a CMZ of $\approx200\pc$ forms, 
very similar to the isothermal case with the same Galactic 
potential adopted in \citet{ridley_etal_17}. We therefore 
regard our isothermal simulations as a first-order approximation to the more 
complicated simulations. We will explore further the effects of these baryonic physics 
on the constraints of the soliton core in follow up studies. 

\subsection{The 3D structure of the CMZ}

We have verified that 3D simulations produce very similar gas flow patterns as in the 2D case. Gas 
cannot go very far above the plane because the deep gravitational potential 
caused by the soliton core and the nuclear bulge. The edge-on $\infty$-shape of 
the CMZ discovered by \citet{molina_etal_11} can also be seen in our 3D simulations. 
However, the $\infty$-shape in the simulations exists only in the first 
$\sim200\Myr$ then gradually decays, as also reported in the SPH simulations of 
\citet{shin_etal_17}. Since it is a transient feature, we prefer not to use 
it as an indicator to constrain the possible existence of the soliton core. 
The observed $\infty$-shape of the CMZ might be caused by a recent 
inflow, as suggested in \citet{sorman_etal_19}. 

%




\section{Summary}
\label{sec:summary}

We have performed high resolution hydrodynamical simulations 
of bar-driven gas flows in a realistic Milky Way potential, 
considering both the effects of a 
baryonic nuclear bulge and a possible dark soliton core. 
The observed size and kinematics of the Central Molecular Zone 
(CMZ) could be reproduced by including 
a compact central mass component, whose mass profile 
is relatively well-constrained by our models. 
If a soliton core is not considered, 
a more compact nuclear bulge than usually 
assumed could match the observed size and kinematics of 
the CMZ. Including a moderate soliton core 
together with a less massive nuclear bulge, 
also nicely agrees with observations. 
An effective sound speed of gas 
around $20\kms$ can help further improve the match. 
The preferred soliton core could be achieved if the 
boson mass is larger than the conventional 
$1.0\times10^{-22}\eV$ by a factor of $2-7$. 
Such a boson mass range is also broadly consistent 
with many other astrophysical constraints. This 
mass range can be further narrowed down by the  
improved determination of the mass-to-light ratio of the nuclear bulge. 
The 3D $\infty$-shape 
structure of the CMZ probably does not offer a tighter constraint on the soliton core 
since it is likely a transient feature that occurs only at the beginning 
of the inflow process. This may imply that the current CMZ 
may have just experienced a recent inflow, which also helps to explain the 
low star formation rate in the CMZ caused by a large 
gas velocity dispersion. 

\acknowledgments 
We would like to thank Scott Tremaine for helpful discussions and 
comments. We thank the anonymous referee for a constructive report, and for 
helping us solving Schr{\"o}dinger-Poisson equation under an external 
potential. We also thank Jonathan Henshaw for 
sharing the NH$_3$ data with us. ZL would like to thank Zhang Jiajun 
for providing comments which have helped to improve the
presentation of the paper. H.S. is grateful to 
Tzihong Chiueh and Tom Broadhurst for stimulating discussions.
The research presented here is partially supported by the National Key 
R\&D Program of China under grant no. 2018YFA0404501, by the National 
Natural Science Foundation of China under grant nos. 11773052, 11333003, 
11761131016, and by a China-Chile joint grant from CASSACA. J.S. 
acknowledges support from an {\it Newton Advanced Fellowship} 
awarded by the Royal Society and the Newton Fund. H.S. 
acknowledges the funding support from the Jade
Mountain Young Scholar Award no. NTU-108V0201, MOST of 
Taiwan under the grant no. 108-2112-M-002-023-MY3,
and the NTU Core Consortium project under the grant no. NTU-CC-108L893401. 
This work made use of the facilities of the Center for High 
Performance Computing at Shanghai Astronomical Observatory. 

\bibliographystyle{aasjournal}
\bibliography{gasdynamics}

\end{document}